\documentclass{appolb}
\usepackage{amssymb,amsmath,bm,bbm}

\usepackage{epsf}
\usepackage{epsfig,rotate}
\usepackage{afterpage}
\usepackage{longtable}
\usepackage{cite}
\usepackage{color}

\newcommand{\vus}{|V_{us}|}
\newcommand{\vcb}{|V_{cb}|}

\newcommand{\vub}{|V_{ub}|}
\newcommand{\vts}{|V_{ts}|}

\def\epe{\varepsilon'/\varepsilon}
\newcommand{\tev}{\, {\rm TeV}}
\newcommand{\gev}{\, {\rm GeV}}

\newcommand{\be}{\begin{equation}}
\newcommand{\ee}{\end{equation}}
\newcommand{\bea}{\begin{eqnarray}}
\newcommand{\eea}{\end{eqnarray}}

\newcommand{\ord}{{\cal O}}

\def\kpn{K^+\rightarrow\pi^+\nu\bar\nu}

\def\klpn{K_{L}\rightarrow\pi^0\nu\bar\nu}

\begin{document}
\title{FCNC Processes Waiting for the Next Decade
\thanks{Presented at the Flavianet topical workshop \it{Low energy constraints on extensions of the Standard Model}}
}
\author{Andrzej J. Buras
\address{Technical University Munich, Physics Department, D-85748 Garching, Germany,\\
 TUM Institute for Advanced Study, D-80333 M\"unchen, Germany \\
   }}
\maketitle
\begin{abstract}
FCNC processes are expected to offer us a deep insight into the physics at 
very short distance scales.
We present a list of 20 goals in quark and lepton flavour physics 
that could be 
reached already in the next decade. This list includes
also flavour conserving observables like 
electric dipole moments of the neutron and leptons and $(g-2)_\mu$. 
Subsequently we will present some aspects of these goals by 
concentrating on supersymmetric flavour models.
A much more extensive presentation of this material can be found in
my recent EPS09 talk \cite{Buras:2009if}.
\end{abstract}

\section{Overture\label{sec:Introduction}}
Flavour-violating and CP-violating processes are very strongly suppressed
and are governed by quantum fluctuations that allow us to probe energy scales 
far beyond the ones explored by the LHC and future colliders. Indeed energy 
scales
as high as $200\tev$ corresponding to short distances in the ballpark of
$10^{-21}{\rm m}$ and even shorter distance scales
can be probed, albeit indirectly, in this manner. Consequently frontiers
in probing ultrashort distance scales belong to flavour physics or more
concretely to very rare processes like particle-antiparticle mixing, rare
decays of mesons and of the top quark, CP violation
and lepton flavour violation. Also electric dipole moments and $(g-2)_\mu$ 
belong to these
frontiers even if they are flavour conserving.
While such tests are not limited by the available energy, they are limited
by the available precision. The latter has to be very high as the 
Standard Model (SM) has been until now very successful and finding 
departures from its predictions has become a real challenge. 

Personally I expect that the coming decade will become the decade 
of discoveries not only at the LHC but in particular in 
high precision flavour experiments like the LHCb, Super-Belle, 
Super Flavour Facility (SFF) in Frascati and dedicated rare $K$ experiments 
around the world. 
Also flavour conserving observables like 
electric dipole moments of the neutron and leptons and $(g-2)_\mu$
will play an important role in this progress.

Recently, I have given a talk at EPS09 in Cracow summarizing the 
present status of flavour theory and presenting flavour
expectations for the coming decade \cite{Buras:2009if}. In view of space limitations I
can discuss here only few points made already in my  EPS09 talk where 
further details can be found. In particular very few references
will be given here. This is, I hope, compensated by  roughly  300 
references given in the EPS09 writeup.

 This presentation consists of three parts. First I will make a list
 of twenty most important goals in this field for the coming decade.
 In the second part I will concentrate on a few 
 topics that I find particularly important and interesting. The third 
 part is dominated by a number of enthusiastic statements that close 
 this report.

\section{Twenty Goals in Flavour Physics for the Next Decade}
We will now list twenty goals in Flavour Physics for the coming decade. 
The order in which these goals will be listed does not represent by any 
means a ranking in importance.

\boldmath
{\bf Goal 1: The CKM Matrix
             from Tree Level Decays}
\unboldmath

This determination would give us the values of the elements of the CKM 
matrix without new physics (NP) pollution. From the present perspective most important are 
the determinations of
$\vub$ and $\gamma$ because they are presently not as well known as
$\vcb$ and $\vus$. However, a precise determination of $\vcb$ is also
important as $\varepsilon_K$, $Br(\kpn)$ and $Br(\klpn)$ are roughly 
proportional to $\vcb^4$. While Super-B facilities accompanied by improved 
theory should be able to determine $\vub$ and $\vcb$ with precision of
$1-2\%$, the best determination of the angle $\gamma$ in the first half
of the next decade will come from the LHCb. An error of a few degrees on 
$\gamma$ should
be achievable around 2015 and this measurement could be further improved 
at Super-B machines.

{\bf Goal 2: Improved Lattice Calculations of Hadronic Parameters}

The knowledge of  meson decay constants $F_{B_s}$, $F_{B_d}$ and of various
$B_i$ parameters with high precision would allow in conjunction with Goal 1
to make precise calculations of the $B_{s,d}$ mixing mass differences 
$\Delta M_{s,d}$, $\varepsilon_K$, 
$Br(B_{s,d}\to\mu^+\mu^-)$ and of other observables in the SM. We  could 
then directly see whether the SM is capable
in describing these observables or not. 
The most recent unquenched calculations allow
for optimism and in fact a very significant progress in the calculation of 
$\hat B_K$, relevant for $\varepsilon_K$, has been made recently. We will discuss its implications in 
Section 3.
\newpage

\boldmath
{\bf Goal 3: Is $\varepsilon_K$ consistent with
 $S_{\psi K_S}$ within the SM?}
\unboldmath

The recent improved value of $\hat B_K$ from unquenched lattice QCD 
acompanied by a more careful look at $\varepsilon_K$ 
\cite{Buras:2008nn} suggests that
the size of CP violation measured in $B_d\to\psi K_S$, might be insufficient
to describe $\varepsilon_K$ within the SM. Clarification of this new tension 
is important as the $\sin 2\beta-\varepsilon_K$ correlation in the SM is presently the
only relation between CP violation in $B_d$ and $K$ systems that can be 
tested experimentally. We will return to this issue in Section 3.

\boldmath
{\bf Goal 4: Is $S_{\psi\phi}$ much larger than its tiny SM value?}
\unboldmath

Within the SM CP violation in the $B_s$ system is predicted to be very small. 
The best known representation of this fact is the value of the mixing 
induced CP asymmetry: $(S_{\psi\phi})_{\rm SM}\approx0.04$. The present data from
CDF and D0 indicate that CP violation in the $B_s$ system could be much
larger $S_{\psi\phi}=0.81^{+0.12}_{-0.32}$. This is a very interesting deviation from the SM. 
Its clarification is of utmost importance and I will return to this question
in Section 3. Fortunately, we should know the answer to this question within the
coming years as CDF, D0, LHCb, ATLAS and CMS will make big efforts to 
measure $S_{\psi\phi}$ precisely.

{\bf Goal 5: Non-Leptonic Two Body B Decays }

The best information on CP violation in the $B$ system to date comes 
from two body non-leptonic decays of $B_d$ and $B^\pm$ mesons. 
The LHCb 
will extend these studies in an important manner
 to $B_s$ and $B_c$ decays. This is clearly
a challenging field not only for experimentalist but in particular for
theorists due to potential hadronic uncertainties. Yet, in the last 
ten years an impressive progress has been made in measuring many 
channels, in particular $B\to\pi\pi$ and $B\to\pi K$ decays, and
developing a number of methods to analyze these data.

I think this field will
continue to be important for the tests of the CKM
framework in view of very many channels whose branching ratios should be
measured in the next decade with a high precision. 
On the other hand in view of potential
hadronic uncertainties present in the branching ratios and direct CP 
asymmetries these observables 
in my opinion will  not provide definite answers about
NP if the latter contributes to them only at the level of $20\%$ or less. 
On the 
other hand mixing induced CP-asymmetries like $S_{\psi K_S}$, 
$S_{\psi\phi}$ and alike being theoretically much cleaner 
will continue to be very important for the tests of NP.

\boldmath
{\bf Goal 6: $Br(B_{s,d}\to\mu^+\mu^-)$}
\unboldmath

In the SM and in several of its extentions  $Br(B_{s}\to\mu^+\mu^-)$ 
is found in the ballpark of $3-5\cdot 10^{-9}$, which is by an order of
magnitude lower than the present bounds from CDF and D0. A discovery of 
$Br(B_{s,d}\to\mu^+\mu^-)$ at $\ord (10^{-8})$ would be a clear signal of
NP, possibly related to Higgs penguins. The LHCb can reach the SM level for 
this branching ratio in the first years of its operation. From my point of 
view, similar to $S_{\psi\phi}$, precise measurements of 
$Br(B_{s}\to\mu^+\mu^-)$ and 
$Br(B_{d}\to\mu^+\mu^-)$ are among the most important goals in flavour 
physics in the coming years. We will discuss both decays in Section 3.

\boldmath
{\bf Goal 7: $B\to X_{s,d}\gamma$, $B\to K^*(\varrho)\gamma$ and 
$A_{\rm CP}^{\rm dir}(b\to s\gamma)$}
\unboldmath

The radiative decays in question, in particular $B\to X_s\gamma$, played 
an important role in constraining NP in the last 15 years because both 
the experimental data and also the theory have been already in a good
shape for some time with the NNLO calculations of $Br(B\to X_s\gamma)$ 
being at the forefront of perturbative QCD calculations in weak decays.
Both theory and experiment reached roughly $10\%$ precision and the 
agreement of the SM with the data is good implying not much 
room left for NP contributions. Still further progress both in theory and 
experiment should be made to further constrain NP models.
Of particular interest is the
direct CP asymmetry $A_{\rm CP}^{\rm dir}(b\to s\gamma)$ that is 
similar to $S_{\psi\phi}$ predicted to be tiny ($0.5\%$) in the 
SM but
could be much larger in some of its extensions.

\boldmath
{\bf Goal 8: $B\to X_sl^+l^-$ and
                     $B\to K^*l^+l^-$}
\unboldmath

While  the branching ratios for $B\to X_sl^+l^-$ and  $B\to K^*l^+l^-$
put already significant constraints on NP, the angular observables, 
CP-conserving ones like the well known forward-backward asymmetry 
and CP-violating ones will definitely be very useful for distinguishing
various extensions of the SM. Recently, a number of detailed analyses 
of various CP averaged symmetries and CP asymmetries provided by the 
angular distributions in the exclusive decay $B\to K^*(\to K\pi)l^+l^-$
have been performed.
In particular the zeroes of some of these 
observables can be accurately predicted. Belle and BaBar provided already
interesting results for the best known forward-backward asymmetry but 
the data have to be improved in order to see whether some sign of NP 
is seen in this asymmetry. Future studies by the LHCb and Super-B machines
will be able to contribute here in a significant manner.

\boldmath
{\bf Goal 9: $B^+\to \tau^+\nu$ and
                     $B^+\to D^0\tau^+\nu$}
\unboldmath

The SM expression for the branching ratio of the tree-level decay $B^+ \to
\tau^+ \nu$ is subject to parametric uncertainties induced  by $F_{B^+}$ and $V_{ub}$.
In order to find the SM prediction for this branching ratio we can 
express them  in terms of $\Delta M_{s,d}$ and $S_{\psi K_S}$, 
all to be taken from experiment.
We then find
\cite{Altmannshofer:2009ne}
\begin{equation}\label{eq:BtaunuSM1}
{Br}(B^+ \to \tau^+ \nu)_{\rm SM}= (0.80 \pm 0.12)\times 10^{-4}.
\end{equation}
 This result
agrees well with a recent result presented by the UTfit collaboration
\cite{Bona:2009cj}. 

On the other hand, the present experimental world avarage based 
on results by BaBar and Belle
reads 
\begin{equation} \label{eq:Btaunu_exp}
{Br}(B^+ \to \tau^+ \nu)_{\rm exp} = (1.73 \pm 0.35) \times 10^{-4}~,
\end{equation}
and is roughly by a factor of 2 higher than the SM value.
We can talk about a tension at the $2.5\sigma$
level.
Interestingly, the tension between  theory and experiment in the 
case of $Br(B^+\to\tau^+\nu)$
 increases in the presence of a tree level $H^\pm$
exchange which interfers destructively with the $W^\pm$ contribution.

The full clarification of a possible
discrepancy between the SM and the data will have to wait for the
data from Super-B machines. Also improved values for $F_B$ from lattice 
and $\vub$ from tree level decays will be important if some NP like
charged Higgs is at work here. The decay $B^+\to D^0\tau^+\nu$ being 
sensitive to different couplings of $H^\pm$ can contribute significantly 
to this discussion but formfactor uncertainties make this decay less
theoretically clean.

\boldmath
{\bf Goal 10: Rare Kaon Decays}
\unboldmath

Among the top highlights of flavour physics in the next decade
will be the measurements of the branching ratios of two {\it golden modes}
$\kpn$ and $\klpn$. $\kpn$ is CP conserving while $\klpn$ is governed by 
CP violation. Both decays are dominated in the SM and many of its
extensions by $Z$ penguin contributions.
It is well known that these decays are theoretically 
very clean and are known in the SM including NNLO QCD corrections and 
electroweak corrections.
Moreover, extensive calculations of isospin breaking effects and 
non-perturbative effects have been done.
The 
present theoretical uncertainties in $Br(\kpn)$ and $Br(\klpn)$ are 
at the level of $2-3\%$ and $1-2\%$, respectively.

Let me stress
that the measurements of the branching ratios in question with an
accuracy of $10\%$ will give us a very important insight into the physics 
at short distance scales. NA62 at CERN in the case of $\kpn$ and KOTO at 
J-PARC in the case of $\klpn$ will tell us how these two decays are
affected by NP.

The decays $K_L\to\pi^0l^+l^-$ are not as 
theoretically clean as the $K\to\pi\nu\bar\nu$ chanels and are less sensitive 
to NP contributions but they probe different operators beyond the SM and 
having accurate branching ratios for them would certainly be very useful.

\boldmath
{\bf Goal 11: $B\to X_s\nu\bar\nu$, $B\to K^*\nu\bar\nu$ and 
        $B\to K\nu\bar\nu$}
\unboldmath

Also  $B$ decays with $\nu\bar\nu$ in the final state provide a very
good test of modified $Z$ penguin contributions, 
but their measurements appear to be
even harder than those of the rare K decays just discussed.

The inclusive decay $B\to X_s\nu\bar\nu$ is theoretically as clean as 
$K\to\pi\nu\bar\nu$ decays but the parametric uncertainties are a bit
larger. The two exclusive channels are affected by  formfactor uncertainties 
but recently in the case of $B\to K^*\nu\bar\nu$ 
and $B\to K\nu\bar\nu$ 
significant progress has been made.
The interesting feature of these three $b\to s\nu\bar\nu$ transitions, in particular when 
taken together, is their sensitivity to right-handed currents. 
Super-B
machines should be able to measure them at a satisfactory level.

\boldmath
{\bf Goal 12: Calculations of Hadronic Matrix Elements in $\epe$}
\unboldmath

One of the important actors of the previous decade in flavour physics was the ratio
$\epe$
 that measures the size of the direct CP
violation in $K_L\to\pi\pi$ 
relative to the indirect CP violation described by $\varepsilon_K$. 
In the SM $\varepsilon^\prime$ is governed by QCD penguins but 
receives also an important destructively interfering
 contribution from electroweak
penguins that is generally much more sensitive to NP than the QCD
penguin contribution.

Here the problem is
the strong cancellation of 
QCD penguin contributions and electroweak penguin contributions to
 $\epe$ and in order to obtain useful predictions  the precision on 
the corresponding hadronic parameters
$B_6$ and $B_8$ should be at least $10\%$. 
Lattice theorists hope to make progress on $B_6$, 
$B_8$ and other $\epe$ related hadronic matrix elements in the coming
decade. 
This would really be good, as the
calculations of  short distance contributions to this ratio (Wilson 
coefficients of QCD and electroweak penguin operators) have been known already 
for 16 years at the NLO level and the  experimental world average  from 
NA48   and KTeV  
$\epe=(16.8\pm 1.4)\cdot 10^{-4}$,
could have an important impact on several extentions of the SM discussed
in the literature if $B_6$ and $B_8$ were known.

\boldmath
{\bf Goal 13: CP Violation in Charm Decays and $D^+(D_s^+)\to l^+\nu$
}
\unboldmath

Charm physics  has been for many years shadowed by the successes of
$K$ decays and $B$ decays, although a number of experimental groups and 
selected theorists have made a considerable
effort to study them. This is due to the GIM mechanism being very effective in
suppressing the FCNC transitions in this sector, long distance contributions
pluguing the evaluation of the $\Delta M_D$ and insensitivity to top
physics in the loops. However, large $D^0-\bar D^0$ mixing discovered
in 2007  
and good prospects for the study of CP violation in these decays at
Super Belle and SFF in Frascati gave a new impetus to this field. 
Also leptonic decays of $D$ mesons remain to be important.

\boldmath
{\bf Goal 14: CP Violation in the Lepton Sector and
                     $\theta_{13}$}
\unboldmath

The mixing angles $\theta_{12}$  and $\theta_{23}$ are already known 
with respectable precision. The obvious next targets in this field
are $\theta_{13}$ and the CP phase $\delta_{\rm PMNS}$.
Clearly the discovery of CP violation in the lepton sector 
would be a very important mile stone
in particle physics for many reasons. In particular the most efficient
explanations of the BAU these days follow from leptogenesis. While in the 
past the
necessary size of CP violation was obtained from new sources of CP
violation at very high see-saw scales, the inclusion of flavour effects,
in particular in  resonant leptogenesis, gave hopes for the explanation
of the BAU using only the phases in the PMNS matrix. This implies certain
conditions for the parameters of this matrix, that is the relevant
$\delta_{\rm PMNS}$, two Majorana phases and $\theta_{13}$. 

\newpage

\boldmath
{\bf Goal 15: Tests of $\mu-e$ and $\mu-\tau$ Universalties}
\unboldmath

Lepton flavour violation (LFV) and the related breakdown of universality 
 can be tested in meson decays by studying the
ratios 
 \be\label{BP-munu}
R_{\mu e}=\frac{Br(K^+\to\mu^+\nu)}
{Br(K^+\to e^+\nu)}, \qquad
R_{\mu\tau}=
\frac{Br(B^+\to\mu^+\nu)}
{Br(B^+\to \tau^+\nu)},
\ee
where the sum over different neutrino flavours
is understood.
The first case is a high precision affair both for experimentalists and 
theorists as both groups decreased the uncertainties in $R_{\mu e}$ 
well below $1\%$  with a precision of $0.5\%$ recently achieved at
CERN. It will continue to constitute an important test of the 
$\mu-e$ universality.
The ratio $R_{\mu\tau}$ is even more sensitive to NP contributions but
it will still take some time before it will be known with good precision.

{\bf Goal: 16 Flavour Violation in Charged Lepton Decays (LFV)}

 The search for  LFV clearly belongs to the most important goals in
 flavour physics.
 In the SM with  
right-handed Dirac neutrinos, the smallness 
 of neutrino masses implies tiny branching ratios for LFV processes.
 For instance
$Br(\mu\to e\gamma)_{\rm SM}\approx 10^{-54}$,
which is more than 40 orders of magnitude below the $90\%$ C.L. upper bound
from the MEGA Collaboration 
\be\label{ueg}
Br(\mu\to e\gamma)< 1.2 \cdot 10^{-11}.
\ee
Therefore any observation of LFV would be a clear sign of NP.
While we hope that new flavoured leptons will be observed at the LHC, 
even if this will not turn out to be the case, LFV has the following virtue:
sensitivity to short distance scales as high as $10^{10}-10^{14}\gev$, in 
particular when the see-saw mechanism is at work.

In order to distinguish various NP scenarios that come close to the
bound in (\ref{ueg}) it will be essential to study a large set
of decays to three leptons in the  final state.
There exist also interesting correlations between leptogenesis and
LFV. Additional correlations relevant for LFV
will be discussed in Section 3.

\boldmath
{\bf Goal 17: Electric Dipole Moments}
\unboldmath

So far CP violation has only been observed in flavour 
violating processes. Non-vanishing electric dipole moments (EDMs) signal 
CP violation in  flavour conserving transitions. 
In the SM CP violation in flavour conserving
processes is very strongly suppressed as best expressed by 
the SM values of electric dipole moments of the neutron and electron
that amount to 
\begin{equation}
d_n\approx 10^{-32}~ {\rm e~ cm.}\qquad d_e\approx 10^{-38}~ {\rm e~ cm.}
\end{equation}  

This should be compared with the present experimental bounds 
\begin{equation}
d_n \le 2.9\cdot  10^{-26}~ {\rm e~ cm.}\qquad 
d_e\le 1.6\cdot10^{-27}~ {\rm e~ cm.}
\end{equation}  
They should be improved in the coming years by 1-2 orders of magnitude.

Similarly to LFV, an observation of a non-vanishing EDM  would 
imply necessarily NP at work. Consequently  correlations between LFV 
and EDMs in specific NP scenarios are to be expected, in particular in 
supersymmetric models, as both types of observables are governed in 
SUSY  by dipole operators. 
We will encounter some examples in Section 3.

\boldmath
{\bf Goal 18: Clarification of the $(g-2)_\mu$ Anomaly}
\unboldmath

The measured anomalous magnetic moment of the electron, $(g-2)_e$, is in 
an excellent agreement with SM expectations. On the other hand, the measured 
 anomalous magnetic moment of the muon, $(g-2)_\mu$, is rougly by 
$3\sigma$  larger than its SM value.
Hadronic contributions to $(g-2)_\mu$ make the comparison of data and theory 
a bit problematic. 
Yet, as this anomaly has been 
with us already for a decade and tremendous effort by a number of theorists
has been made to clarify this issue, this anomaly could indeed come from 
NP. 

The MSSM with large $\tan\beta$ and sleptons with masses below $400\gev$ 
is capable of reproduce the experimental value of $(g-2)_\mu$ 
 provided the $\mu$ parameter
in the Higgs Lagrangian has a specific sign. At SFF  also
$(g-2)_\tau$ can be measured and it is also sensitive to NP contributions.

{\bf Goal 19: Flavour Violation at High Energy}

Our presentation deals mainly with tests of flavour and CP violation in
low energy processes. However, at the LHC it will be possible to investigate
 these phenomena also in high energy processes, in particular in top quark
decays.

{\bf Goal 20: Construction of a New Standard Model (NSM)}

Finally, in view of so many parameters present in basically all extensions
of the SM like the MSSM, the LHT model and RS models, it is unlikely from my point of view 
that any of the models
studied presently in the literature will turn out to be the new
model of elementary particle physics. On the other hand various structures, 
concepts
and ideas explored these days in the context of specific models may well turn
out to be included in the NSM that is predictive, consistent with all the 
data and giving explanation of observed hierarchies in fermion masses and
mixing matrices. While these statements may appear to be very naive, it is
a fact that the construction of the NSM is the main goal of elementary
particle physics
and every theorist, even as old as I am,
 has a dream that the future NSM will carry her (his) name.

\section{Waiting for Signals of New Physics in FCNC Processes}

\subsection{Strategies for the Search for New Physics in the Next Decade}
Let us first emphasize that until now only $\Delta F=2$ FCNC processes 
 could be used in the UTfits. The measured $B\to X_s\gamma$ and $B\to X_s
 l^+l^-$ decays  and their exclusive 
counterparts are sensitive to $\vts$ that has nothing to do with the usual
UT plots. The same applies to the observables in the $B_s$ system, which
with the $S_{\psi\phi}$ anomaly observed by CDF and D0 and the studies 
of rare $B_s$ decays at the Tevatron and later at  the LHC are becoming central
for flavour physics. Obviously these comments also apply to all lepton 
flavour violating processes.

In this context 
a special role is played by $Br(\kpn)$ and $Br(\klpn)$ as their values
allow a theoretically clean construction of the UT in a manner complementary 
to its present determinations: the height of the UT 
is determined from $Br(\klpn)$ and the side $R_t$ from $Br(\kpn)$. Thus 
projecting the results of future experimental results for these two 
branching ratios on the $(\bar\varrho,\bar\eta)$ plane could 
be a very good test of the SM.

Yet, generally 
I do not think that in the 
context of the search for the NSM (see Goal 20) it is a good strategy  to
project the results of all future measurements of rare decays 
on the $(\bar\varrho,\bar\eta)$ plane or any other of five planes 
related to the remaining unitarity triangles. This would only teach us about
possible inconsistences within the SM but would not point towards 
a particular NP model.

In view of this, here comes a proposal for the strategy for searching 
for NP in the next decade, in which hopefully the side $R_b$ and the angle 
$\gamma$ in the UT 
will
be precisely measured, CP violation in the $B_s$ system explored and 
many goals listed in the previous section reached.
This strategy proceeds in three steps:

{\bf Step 1}

In order to study transparently possible tensions between 
$\varepsilon_K$, $\sin 2\beta$, $\vub$, $\gamma$ and $R_t$ let us 
leave the $(\bar\varrho,\bar\eta)$ plane and {go to}
the $R_b-\gamma$ plane \cite{Buras:2002yj} suggested 
already several years ago and recently strongly supported by the 
analysis in 
\cite{Altmannshofer:2007cs,Altmannshofer:2009ne}. 
The $R_b-\gamma$ plane is shown in Fig.~\ref{fig:UTfit}. We will explain this 
figure in the next subsection.

{\bf Step 2}

In order to search for NP in rare $K$, $B_d$, $B_s$, $D$ decays, in 
CP violation in $B_s$ and charm decays, in LFV decays, in EDMs and 
$(g-2)_\mu$ let us go to specific plots that exhibit correlations 
between various observables. As we will see below such correlations 
will be crucial to 
distinguish various NP scenarios.
Of particular importance are the correlations between the 
CP asymmetry $S_{\psi\phi}$ and
$B_s\rightarrow\mu^+\mu^-$, between 
the anomalies in $S_{\phi K_s}$ and $S_{\psi\phi}$, 
between $\kpn$ and $\klpn$, between $\kpn$ and $S_{\psi\phi}$,
between $S_{\phi K_s}$ and $d_e$, between $S_{\psi\phi}$ and $(g-2)_{\mu}$ and
also those involving lepton flavour violating decays.

{\bf Step 3}

In order to monitor the progress made in the next decade when additional 
data on flavour changing processes will become available, it is useful to
construct a ``DNA-Flavour Test'' of NP models \cite{Altmannshofer:2009ne} 
including Supersymmetry, the LHT
model, the RS models and various supersymmetric flavour models and other 
models,
with the aim to distinguish between these NP scenarios in a global manner.

Having this strategy in mind we will in the rest of this writing illustrate
Steps 1 and 2  on several examples. The full table representing 
``DNA-Flavour Test'' can be found in \cite{Altmannshofer:2009ne}. Here we
will reduce the illustration of Step 3 to few observables.

\boldmath
\subsection{Tension in the $R_b-\gamma$ Plane}
\unboldmath
Recently, in connection with Goal 3 some tensions in the UTFits 
have been identified
in~\cite{Lunghi:2008aa,Buras:2008nn}. 

\begin{figure}[t]
\includegraphics[width=1.\textwidth]{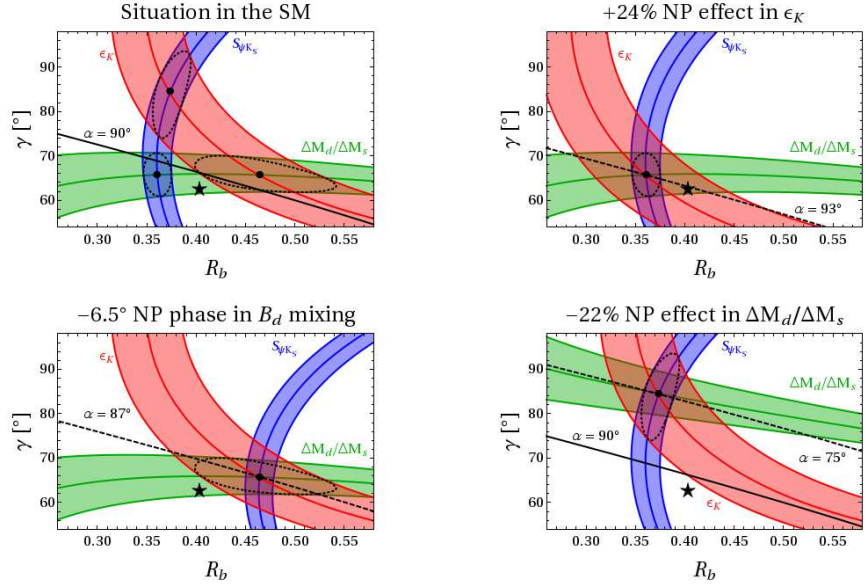}
\caption{\small
The $R_b-\gamma$ plane as discussed in the text. For further explanations see
\cite{Altmannshofer:2009ne}}
\label{fig:UTfit}
\end{figure}

In order to see them transparently 
let us have now a look at the $R_b-\gamma$ plane in 
Fig.~\ref{fig:UTfit} taken from \cite{Altmannshofer:2009ne}. 
There, in the upper left plot the {\it blue} ({\it green})
region corresponds to the 1$\sigma$ allowed range for $\sin2\beta$ ($R_t$) as
calculated in the SM. The {\it red} region
corresponds to $|\epsilon_K|$ in the SM. Finally the solid black line
corresponds to $\alpha=90^\circ$ that is close to the value favoured by UT
fits and other anylyses.

It is evident that there is a tension between various regions as there are
three different values of $(R_b, \gamma)$, dependending on 
which two constraints
are simultaneously applied. The four immediate solutions to this tension
are as follows:

{\bf 1.}  There is a positive NP effect in $\epsilon_K$ 
while $\sin 2\beta$ and $\Delta M_d/\Delta M_s$ are SM-like~
\cite{Buras:2008nn}, as shown by the upper
right plot of Fig.~\ref{fig:UTfit}.
The required effect in $\epsilon_K$ could be for instance achieved within
models with CMFV by a positive shift in the relevant one-loop box diagram 
function.
Alternatively, new non-minimal sources of flavour violation
relevant only for the $K$ system could solve the problem.
Note that this solution corresponds to $\gamma \simeq 66^\circ$, 
$R_b \simeq 0.36$ and  $\alpha \simeq 93^\circ$ in accordance with the usual UT analysis.

{\bf 2.} $\epsilon_K$ and $\Delta M_d/\Delta M_s$ are  NP free while $S_{\psi
  K_S}$ is affected by a NP phase $\phi_{B_d}$ in 
$B_d$ mixing of approximately 
$-7^\circ$. This is shown in the lower left plot of Fig.~\ref{fig:UTfit}.
The predicted value for $\sin2\beta$ is now shifted to 
$\sin 2\beta\approx 0.85$ 
\cite{Lunghi:2008aa,Buras:2008nn}. This value is
  significantly larger than the measured $S_{\psi K_S}$ which allows to fit
  the experimental value of $\epsilon_K$.
Note that this solution is characterized by a large value of $R_b \simeq 0.47$, that is significantly larger than its exclusive determinations but still compatible with the inclusive determinations. The angles $\gamma \simeq 66^\circ$ and $\alpha \simeq 87^\circ$ agree with the usual UT analysis.

{\bf 3. } $\epsilon_K$ and $S_{\psi K_S}$ are NP free while the determination
of $R_t$ through $\Delta M_d/\Delta M_s$ is affected by NP. 
This is shown in the lower 
right plot of Fig.~\ref{fig:UTfit}. 
In that scenario one finds  $\Delta M_d^{\rm SM} / \Delta M_s^{\rm SM}$ 
to be much higher than the actual measurement. In order to agree exactly with
the experimental central value, one needs a NP contribution to $\Delta M_d /
\Delta M_s$ at the level of $-22\%$ leading to an 
increased value of $R_t$ that compensates the negative effect of NP in
$\Delta M_d/\Delta M_s$. This in turn  allows to fit the experimental value 
of $\epsilon_K$.
This solution is characterized by a large value of $\gamma\simeq 84^\circ$ and $\alpha$
much below $90^\circ$. The latter fact could become problematic for this solution when the determination
of $\alpha$ further improves.

 {\bf 4.} The value of $\vcb$ is significantly increased to roughly 
           $43.5\cdot 10^{-3}$, which seems rather unlikely.

The first  three NP scenarios characterized by black points in
Fig.~\ref{fig:UTfit} will be clearly distinguished from each other once the
values of $\gamma$ and $R_b$ from tree level decays will be precisely
known. Moreover, if  future measurements of $(R_b,\gamma)$ will select a
point in the $R_b - \gamma$ plane that differs from the black points in
Fig.~\ref{fig:UTfit}, it is likely that NP will simultaneously enter
$\epsilon_K$, $S_{\psi K_S}$ and $\Delta M_d / \Delta M_s$. It will be
interesting to monitor future progress in the $R_b-\gamma$ plane.

\subsection{Correlations in Supersymmetric Flavour Models}
The correlations between various observables in the LHT model 
\cite{Blanke:2009am} and in 
the RS model with custodial protection 
\cite{Blanke:2008zb,Blanke:2008yr} 
have been already discussed at
this workshop by Recksiegel and Duling, respectively. Therefore I will
confine my discussion to
supersymmetric flavour
models (SF) with flavour symmetries that allow a simultaneous understanding of 
the flavour structures in the Yukawa couplings and in SUSY soft-breaking 
terms, adequately suppressing FCNC and CP-violating phenomena and solving
SUSY flavour and CP problems. A recent detailed study of various SF models 
has been performed in \cite{Altmannshofer:2009ne} and I will summarize the 
results of this work here. 

We have analysed 
the following representative scenarios in which NP contributions are characterized
by:
\begin{itemize}
\item [i)] The dominance of right-handed (RH) currents 
(abelian model by Agashe and Carone),
\item [ii)] Comparable left- and right-handed currents with CKM-like mixing
  angles represented by the special version (RVV2) 
of the non-abelian $SU(3)$ 
model by
Ross, Velasco and Vives as discussed recently by Calibbi et al and 
the model by Antusch, King and Malinsky (AKM),
\item [iii)] The dominance of left-handed (LH) currents in non-abelian 
models ($\delta$LL) .
\end{itemize}

\begin{figure}[tbp]
\begin{center}
\includegraphics[width=2.45in]{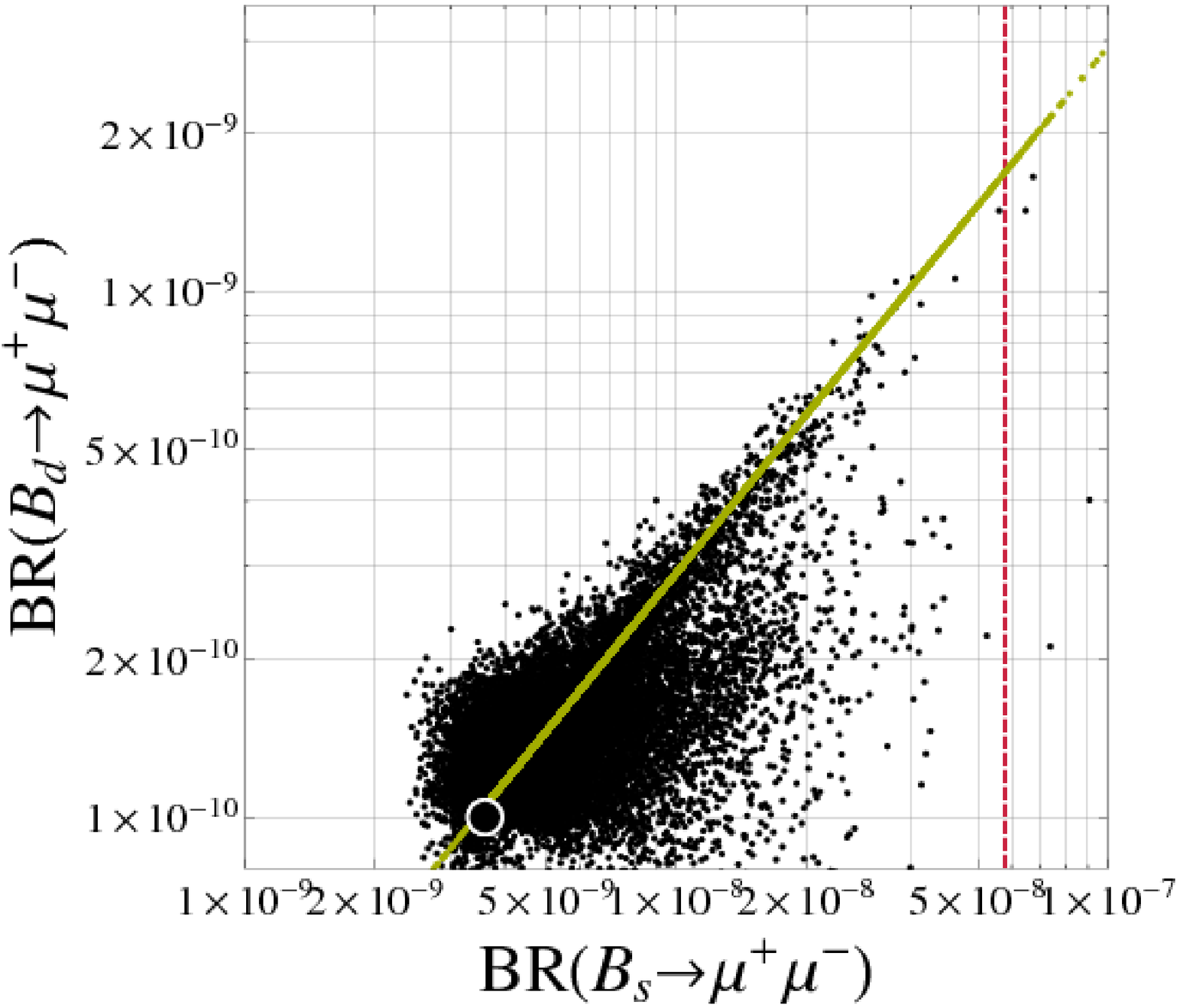}
\includegraphics[width=2.45in]{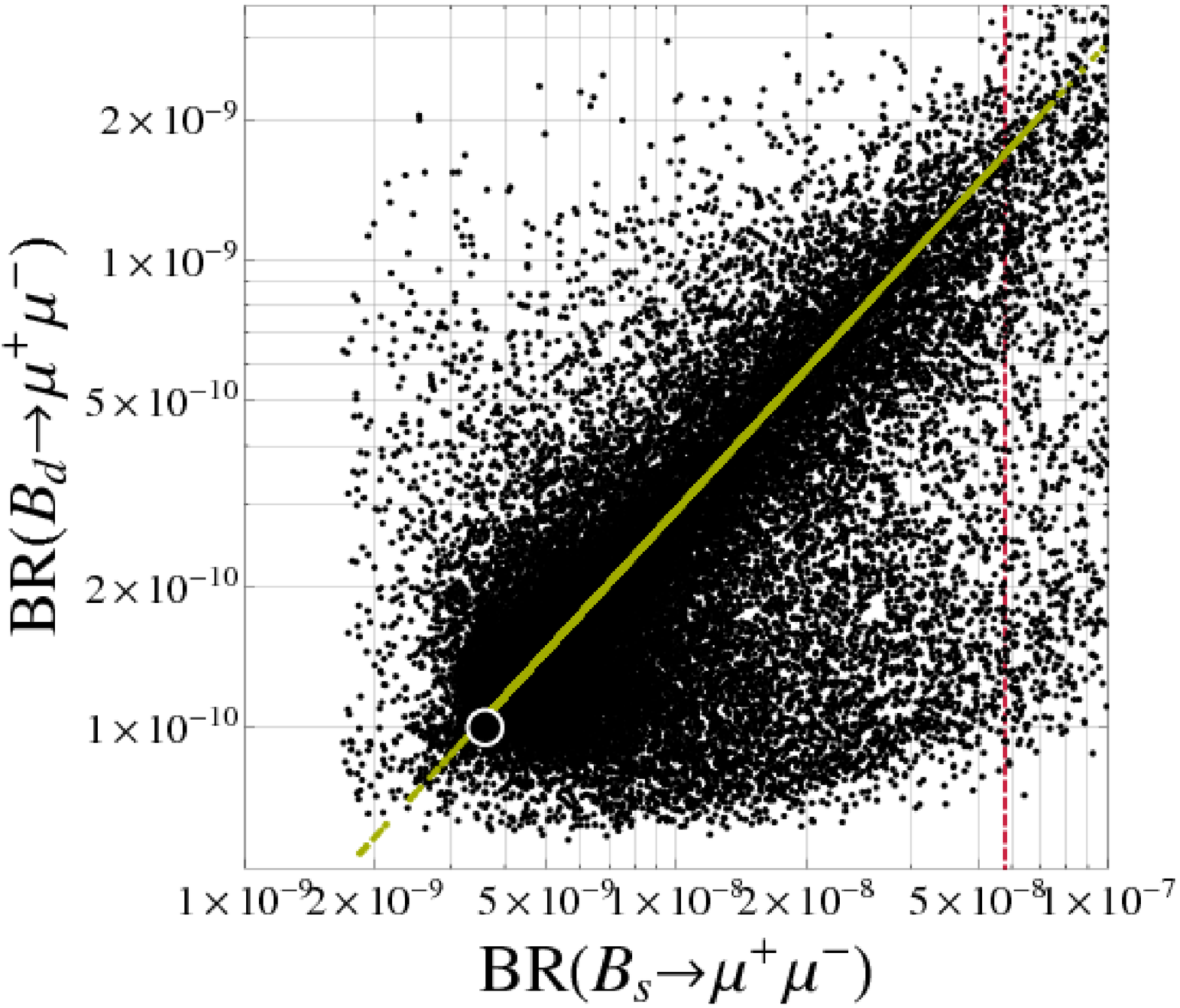}
\end{center}
\caption{\label{Bds} $B_{d,s}\to \mu^+\mu^-$ branching ratios in the
RVV2 model (left) and the $\delta$LL model (right) as obtained 
in \cite{Altmannshofer:2009ne}. }
\end{figure}

 We find \cite{Altmannshofer:2009ne}: 

{\bf 1.}
The ratio 
$Br(B_d\to\mu^+\mu^-)/Br(B_s\to\mu^+\mu^-)$ in the AC and RVV2  models is 
dominantly below its CMFV prediction and can be much 
smaller than the latter.
In the AKM model this ratio stays much closer to the MFV value of roughly 
$1/33$ and can be smaller or larger than 
this value with equal probability.
Still, values of $Br(B_d\to\mu^+\mu^-)$ as high as $1\times 10^{-9}$ are
possible in all these models as $Br(B_s\to\mu^+\mu^-)$ can be strongly 
enhanced. We show this in the case of the RVV2 model in the left plot of
Fig.~\ref{Bds}.

{\bf 2.}
Interestingly,  in the $\delta$LL-models, the ratio 
$Br(B_d\to\mu^+\mu^-)/Br(B_s\to\mu^+\mu^-)$ can not only deviate
significantly from its CMFV value, but in contrast to the models with 
right-handed currents considered by us can also be larger that the MFV value. 
Consequently,
$Br(B_d\to\mu^+\mu^-)$ as high as $(1-2)\times 10^{-9}$ is  
possible while being consistent with the bounds on all other observables,
in particular the one on $Br(B_s\to\mu^+\mu^-)$. We show this in the 
right plot of Fig.~\ref{Bds}.

\begin{figure}[thbp]
\begin{center}
\includegraphics[width=2.45in]{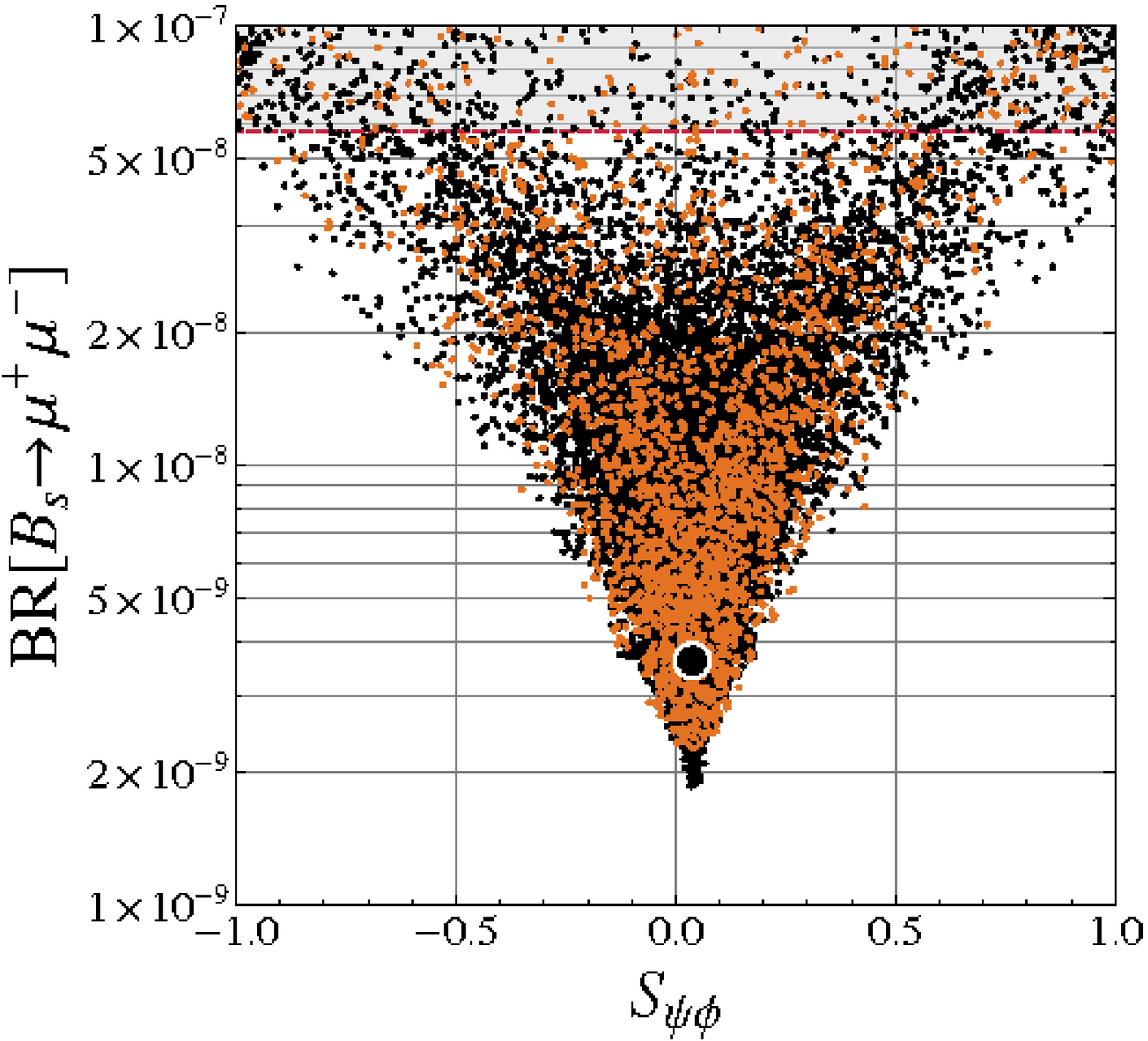}
\includegraphics[width=2.45in]{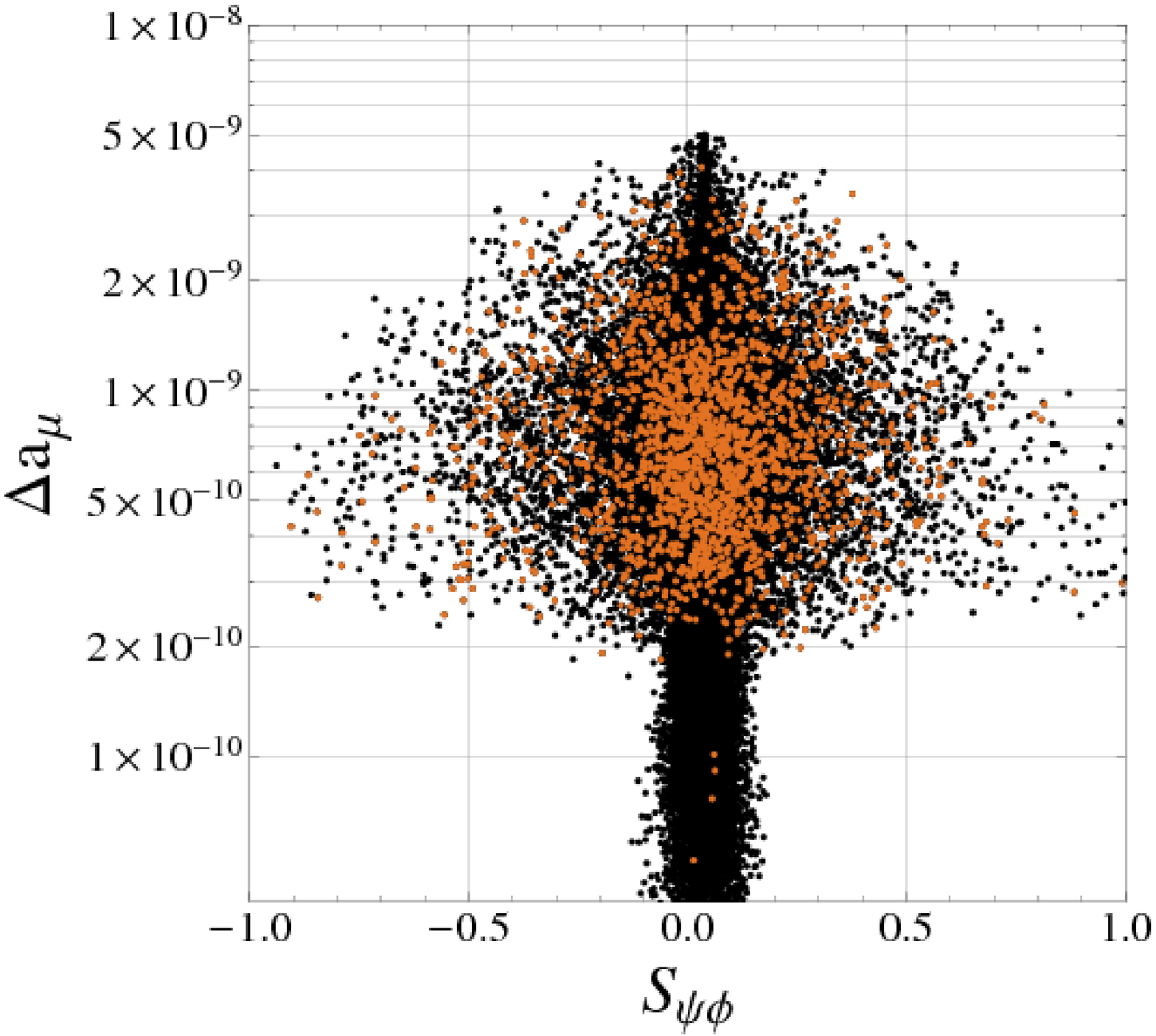}
\end{center}
\caption{\label{AC} $Br(B_{s}\to \mu^+\mu^-)$ vs. $S_{\psi\phi}$ (left)
and $\Delta a_\mu$ vs. $S_{\psi\phi}$ (right) in the AC model as obtained 
in \cite{Altmannshofer:2009ne}. $a_\mu=(g-2)_\mu/2$.}
\end{figure}

{\bf 3.}
The $S_{\psi\phi}$ anomaly within 
the supersymmetric flavour models with
right-handed  currents implies, in the case of the AC and  AKM models,
values of
$Br(B_s\to\mu^+\mu^-)$ as high as several $10^{-8}$. This are 
very exciting news 
for the CDF, D0 and LHCb experiments! In the 
RVV2 model such values are also possible but not necessarily implied
by the large value of $S_{\psi\phi}$. We show one example of this 
spectacular correlation for the case of the AC model in the left plot 
of Fig.~\ref{AC}. 

\begin{figure}[thbp]
\begin{center}
\includegraphics[width=2.45in]{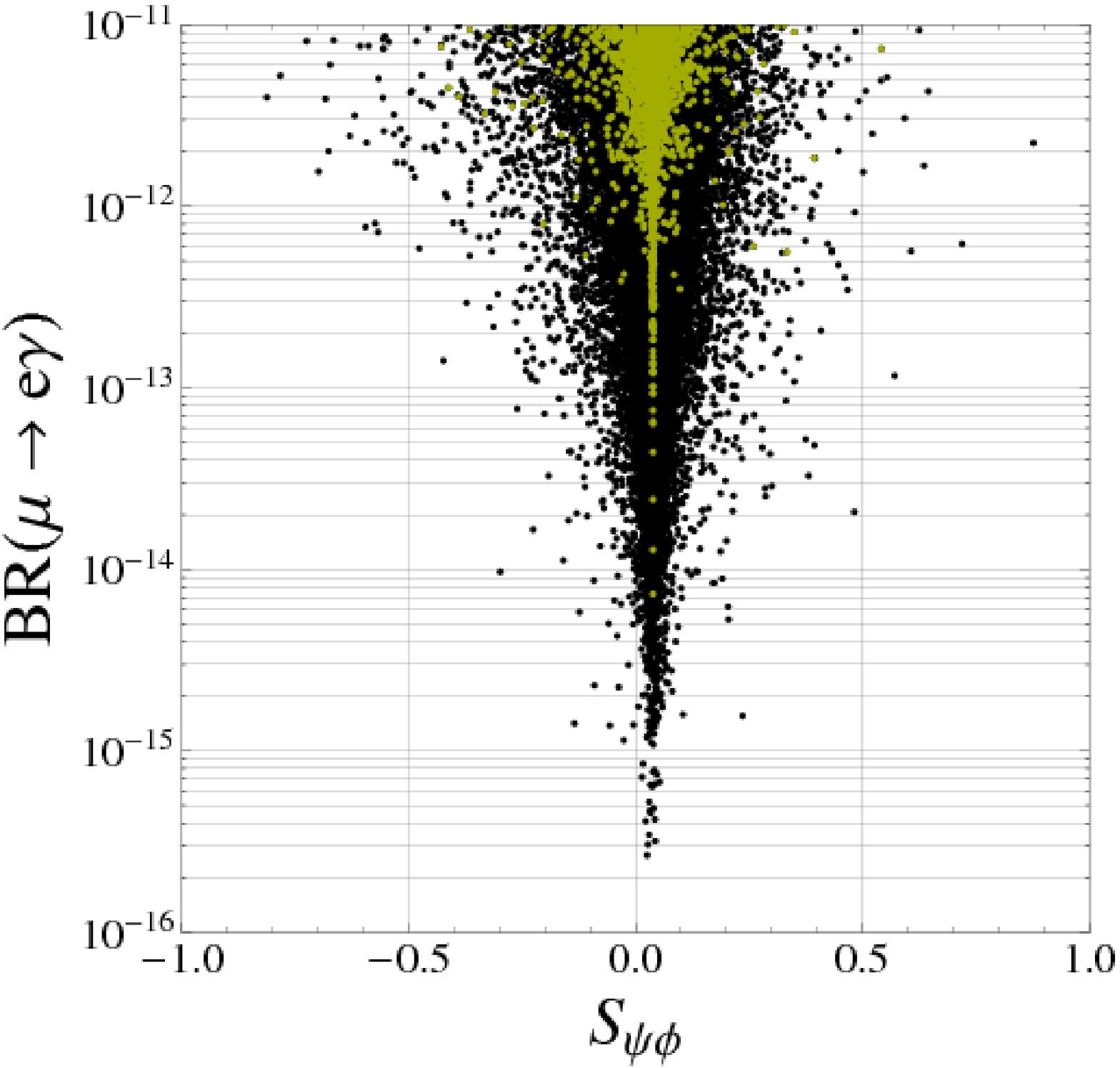}
\includegraphics[width=2.45in]{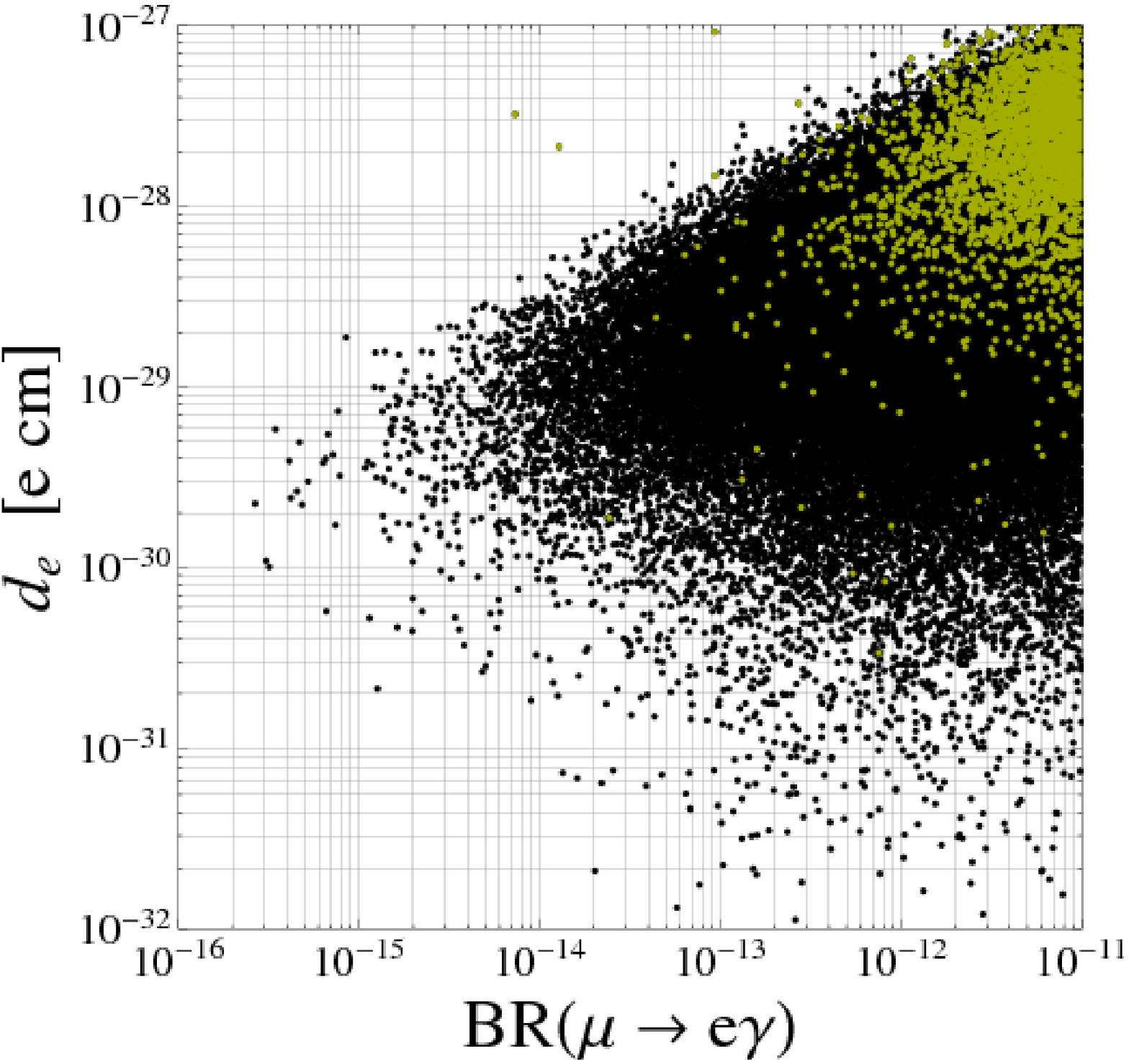}
\end{center}
\caption{\label{RVV2} $Br(\mu\to e\gamma)$ vs. $S_{\psi\phi}$ (left)
and $d_e$ vs. $Br(\mu\to e\gamma)$ (right) in the RVV2 model as obtained 
in \cite{Altmannshofer:2009ne}. The green points explain the $(g-2)_\mu$
anomaly at $95\%$ C.L., i.e. $\Delta a_\mu\ge 1\times 10^{-9}$. }
\end{figure}

{\bf 4.} In the AC model a large value of $S_{\psi\phi}$ implies a solution 
to the $(g-2)_\mu$ anomaly as seen in the right plot of Fig.~\ref{AC}.
In the RVV2 and the AKM models additionally 
$Br(\mu\to e\gamma)$ in 
the reach of the MEG experiment is implied.  In the case of the RVV2 model,
$d_e\ge 10^{-29}$ e cm. is predicted, while in the AKM model it is typically
smaller.
 Moreover, in the case of the RVV2 model, 
$Br(\tau\to\mu\gamma)\ge 10^{-9}$ is then
 in the reach of Super-B machines, while this is not the case in the AKM model.
 Some of these results are illustrated in Fig.~\ref{RVV2}.

\begin{figure}[htbp]
\begin{center}
\includegraphics[width=2.45in]{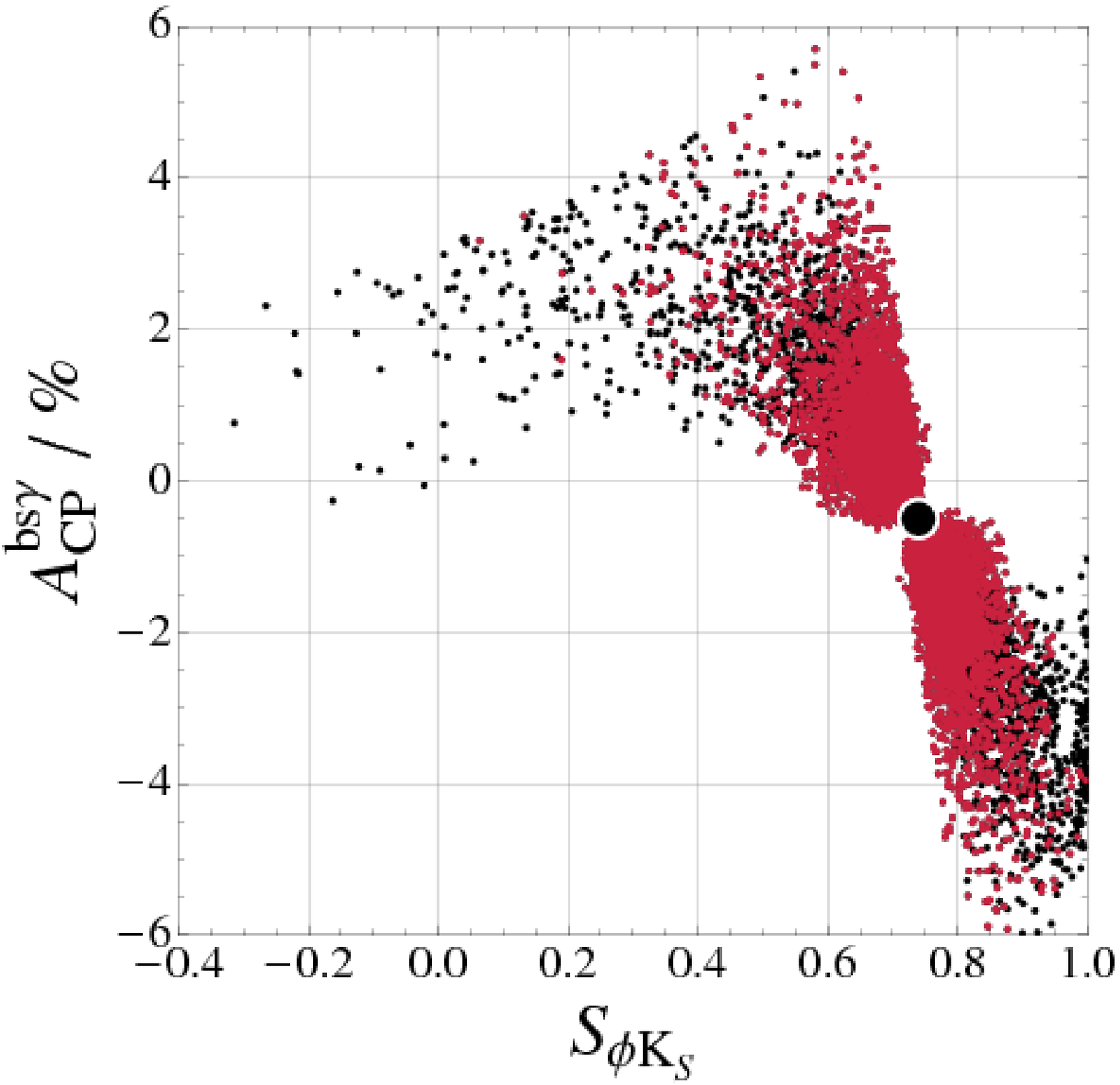}
\includegraphics[width=2.45in]{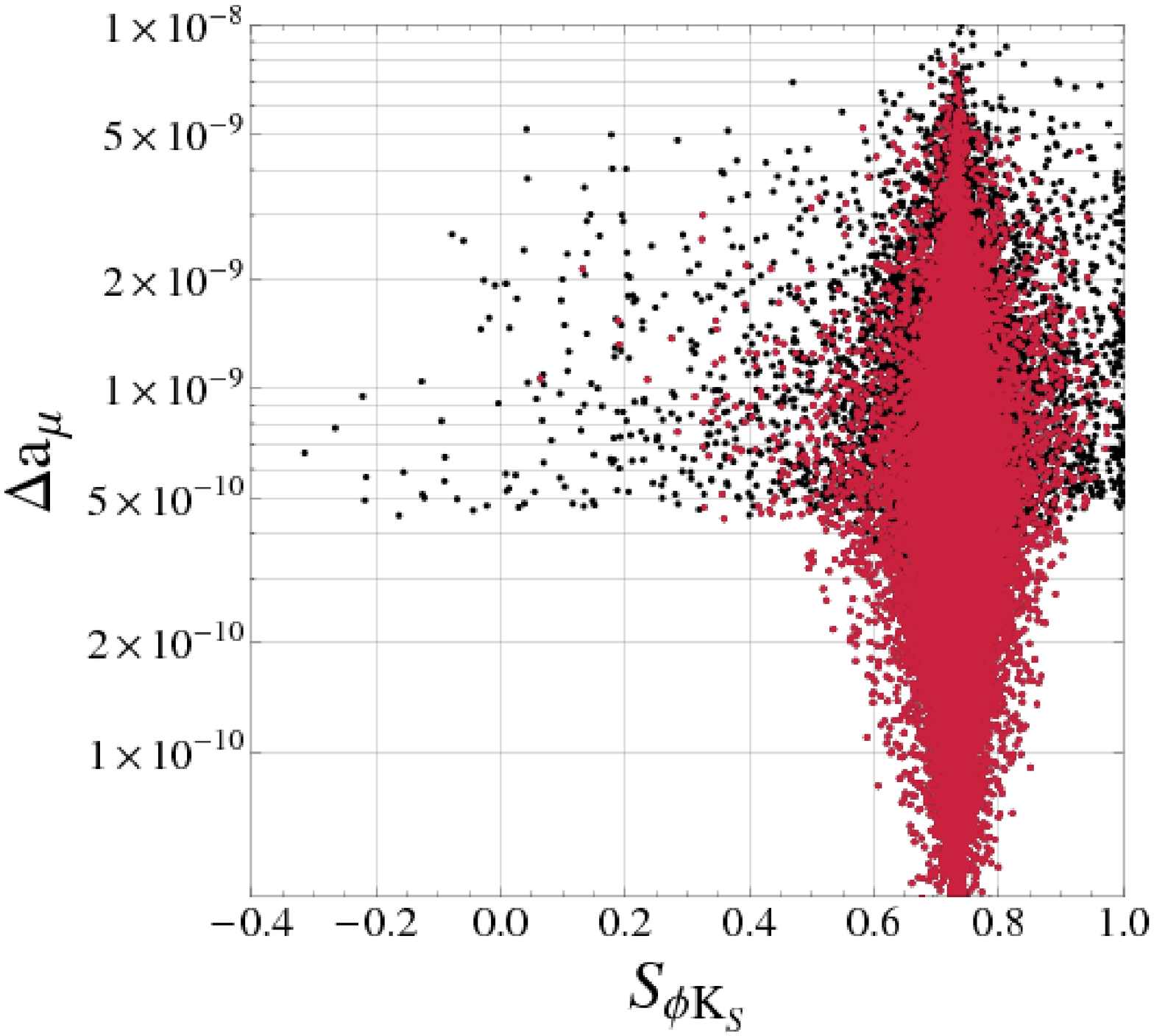}
\end{center}
\caption{\label{dLL} $A_{\rm CP}^{bs\gamma}$ vs. $S_{\phi K_S}$ (left)
and $\Delta a_\mu$ vs. $S_{\phi K_S}$ (right) in the $\delta$LL model as 
obtained 
in \cite{Altmannshofer:2009ne}. The red points satisfy 
$Br(B_s\to\mu^+\mu^-)\le 6 \times 10^{-9}$.}
\end{figure}

{\bf 5.}
In the supersymmetric models with exclusively left-handed currents ($\delta$LL), 
the desire to explain
the $S_{\phi K_S}$ anomaly implies automatically a solution to the 
$(g-2)_\mu$ anomaly and the direct CP asymmetry in $b\to s\gamma$ much
larger than its SM value. We illustrate this in Fig.~\ref{dLL}.
This is in contrast to the models with right-handed  currents
where the $A_{\rm CP}^{bs\gamma}$ remains SM-like.

\subsection{Maximal Enhancements in Various Models}
 Finally, we present in Table~\ref{tab:summary} approximate maximal enhancements for
$Br(\kpn)$, $Br(\klpn)$, $Br(B_s\to\mu^+\mu^-)$ and $S_{\psi\phi}$
in various models. Brief description of all these models can be found
in \cite{Buras:2009if}.

{
\begin{table}[ht]
\renewcommand{\arraystretch}{1}\setlength{\arraycolsep}{1pt}
\center{\begin{tabular}{|c|c|c|c|c|}
\hline
Model & $Br(\kpn)$ & $Br(\klpn)$ & $Br(B_s\to\mu^+\mu^-)$ &
$S_{\psi\phi}$\\
\hline
 CMFV  & $20\%$ & $20\%$ & $20\%$ & 0.04 \\
 MFV  & $30\%$ & $30\%$ & $1000\%$ & 0.04 \\
 AC  & $2\%$ & $2\%$ & $1000\%$ & 1.0 \\
 RVV2  & $10\%$ & $10\%$ & $1000\%$ & 0.50 \\
 AKM  & $10\%$ & $10\%$ & $1000\%$ & 0.30 \\
$\delta$LL  & $2\%$ & $2\%$ & $1000\%$ & 0.04 \\
LHT  & $150\%$ & $200\%$ & $30\%$ & 0.30 \\
RSc  & $60\%$ & $150\%$ & $10\%$ & 0.75 \\
\hline 
\end{tabular}  }
\caption {Approximate maximal enhancements for various observables in different 
models of NP. In the case of $S_{\psi\phi}$ we give 
the maximal positive values. The NP models have been  defined in \cite{Altmannshofer:2009ne}.}
\label{tab:summary}
\renewcommand{\arraystretch}{1.0}
\end{table}
}

\section{Final Messages and Five  Big Questions}

 In our search for a more fundamental theory we need to improve our 
 understanding of  flavour physics.
 The study of flavour physics in conjuction with direct collider searches
 for new physics, with electroweak precision tests and cosmological
 investigations will result one day  in a NSM. When this will
 happen is not clear at present. Afterall,
 35 years have passed since the completion of the present SM and no fully
 convincing  candidate for the NSM exists in the literature. On the
 other hand in view of presently running and upcoming experiments, 
the next decade
 could be like 1970's in which practically every year a new important
 discovery has been made. Even if by 2019 a NSM may not exist yet, it
 is conceivable that we will be able to answer the following crucial
 questions by then:
\begin{itemize}
\item
 Are there any fundamental scalars with masses $M_s\le 1\tev$?
\item
 Are there any new fundamental fermions like vector-like fermions or
 the 4th generation of quarks and leptons?
\item
 Are there any new gauge bosons leading to new forces at very short
 distance scales and an extended gauge group?
\item
 What are the precise patterns of interactions between the gauge bosons,
 fermions and scalars with respect to flavour and CP Violation?
\item
 Can the answers to these four questions help us in understanding the BAU
 and other fundamental cosmological  questions?
\end{itemize}

 There are of course many other profound questions in 
  elementary particle physics and cosmology but from my point of view I
 would really be happy if in 2019 satisfactory answers to the five
 questions posed above were available.
 
 In this presentation I wanted to emphasize that
 many observables in the quark and lepton flavour sectors have not 
 been measured yet or only poorly measured and that flavour physics 
 only now enters the precision era. Indeed, spectacular 
 deviations from the SM and MFV expectations are still possible in 
 flavour physics.
 The interplay of the expected deviations with direct searches at
 Tevatron, LHC and later at ILC will be most interesting.
 Finally,
 the correlations between various 
 observables will pave the road to the NSM.

{\bf Acknowledgements}\\
I would like to thank the organizers 
 for inviting me to this very pleasent and interesting workshop.
I would  also like to thank all my collaborators for a wonderful
time we spent together exploring different avenues beyond the Standard
Model. 
This research was supported by the Deutsche
Forschungsgemeinschaft (DFG) under contract BU 706/2-1, the DFG Cluster of
Excellence `Origin and Structure of the Universe' and by the German
Bundesministerium f{\"u}r Bildung und Forschung under contract 05HT6WOA.


\end{document}